# Topological anomalies in the off-diagonal Ehrenfest theorem and their role on optical transitions in solar cells

Georgios Konstantinou[1], Konstantinos Moulopoulos[2]

[1]FOSS Research Centre for Sustainable Energy, PV Technology, University of Cyprus, 75 Kallipoleos Str., Nicosia, 1678, Cyprus

[2]Department of Physics, University of Cyprus, 75 Kallipoleos Str., Nicosia, 1678, Cyprus

We analytically demonstrate the emergence of surface non-Hermitian boundary contributions that appear in an extended form of the quantum Ehrenfest theorem and are crucial (although so far overlooked) in the calculation of optical matrix elements that govern the optical transitions in semiconductors, e.g. solar cells. Their inevitable existence, strongly related to the boundary conditions of a given quantum mechanical problem, is far-reaching in the sense that they play a crucial role in the dynamics of solar absorption and the corresponding optical transitions that follow. Processes like optical transitions in localized and delocalized states and probabilities of intermolecular transitions can be investigated through this generalized off-diagonal Ehrenfest theorem, employed in the present work for the first time. An explicit demonstration of bulk-boundary correspondence is shown, as the extended Ehrenfest theorem can be separated into bulk and surface contributions, each behaving separately from the other, but at the end collaborating to give the correct time-derivative of the desired optical element; this paves the way for future application of the extended theorem to optical transitions in topologically nontrivial quantum systems. It is also demonstrated through two examples in the literature as well as through a new example (of a system exhibiting the Integer Quantum Hall Effect) that non-Hermitian boundary terms (that have been designated 'topological anomalies' in the mathematical literature) may be expected to be quantized, especially in topologically nontrivial quantum systems but also in certain conventional ones.

1. ## Introduction

The well-known Ehrenfest theorem of Quantum Mechanics describes the time-flow of the mean value of a vector operator $\vec{B}$ through the following elegant relation (the also well-known Heisenberg equation):

$$\frac{d}{dt}\langle\Psi|\vec{B}|\Psi\rangle = \left\langle\Psi\left|\frac{\partial\vec{B}}{\partial t}\right|\Psi\right\rangle + \frac{i}{\hbar}\langle\Psi|[H,\vec{B}]|\Psi\rangle \quad (1)$$

where $|\Psi\rangle$ is **any** state $|\Psi(t)\rangle$ of the system, solution of the *t*-dependent Schrödinger equation, and $[H,\vec{B}]$ denotes the commutator of $\vec{B}$ with $H$. The above, if viewed as a continuity equation, states that the operator $\vec{B}$ **is conserved** *(its mean value is independent of time)* if either $\vec{B}$ is time-independent and commutes with $H$, or whenever $\frac{\partial\vec{B}}{\partial t} = -\frac{i}{\hbar}[H,\vec{B}]$, i.e. in the case that $\vec{B}$ is an invariant operator [3]. This statement is not however generally true (in the sense that a local form of the above theorem may lack a divergence of a current density); indeed it has been explicitly proved in [1] that the following generalized Ehrenfest theorem is valid (with $B_l$ a certain Cartesian component of the vector operator $\vec{B}$, with *l=1,2,3*):

$$\frac{d}{dt}\langle\Psi|B_l|\Psi\rangle = \left\langle\Psi\left|\frac{\partial}{\partial t}B_l\right|\Psi\right\rangle + \frac{i}{\hbar}\langle\Psi|[H,B_l]|\Psi\rangle - \oint \vec{J}_{gen}\cdot d\vec{S} \quad (2),$$

with $\vec{J}_{gen} = \frac{i\hbar}{2m}\left(\vec{\nabla}\Psi^* B_l\Psi - \Psi^*\vec{\nabla}(B_l\Psi)\right) - \frac{q}{mc}\vec{A}\Psi^* B_l\Psi$ **(3)** the generalized current density of the quantity $B_l$, and $\vec{A}$ is any magnetic vector potential present in the system. The last flux term in eq. (2) across the system boundaries describes non-Hermitian effects that are emergent (and are strictly resulting from the boundary conditions). Although we have used notation for the surface flux of $\vec{J}_{gen}$ (proper for a 3D system), eq. (2) is also valid for any dimensionality (with the last term being a line integral of the component of $\vec{J}_{gen}$ perpendicular to the displacement element $d\vec{l}$ for 2D systems), or the difference of values of $\vec{J}_{gen}$ between two points (the ends of a 1D system). If we also define $\rho_{gen} = \Psi^* B_l \Psi$ **(4)** a generalized density of $B_l$, then eq. (2) can also be written in differential form, namely:

$$\vec{\nabla}\cdot\vec{J}_{gen} + \frac{\partial \rho_{gen}}{\partial t} = \Psi^*\left(\frac{\partial}{\partial t}B_l + \frac{i}{\hbar}[H,B_l]\right)\Psi \quad (5)$$

Eq. (5) is actually a generalized local conservation law: the local contribution to $\langle\Psi|B_l|\Psi\rangle$ together with its flow, satisfy a continuity balance only if the source term that appears on the right hand side of (5) vanishes. The above extended form of the Ehrenfest theorem is more complete and has potential consequences on many elementary Quantum Mechanical problems (see Ref. [1] for a few examples). Also, it resolves some previously noticed quantum mechanical paradoxes (see Ref. [4], as well as Ref. [5] for an observation on the Hypervirial theorem). The non-Hermitian boundary contributions that are contained in the extended theorem have been characterized by mathematicians as "topological anomalies" (see i.e. eq.(7) of [11]) and they originate from (i.e. they are non-vanishing in) cases that the input operator $B_l$ is a "bad operator" for the given boundary conditions (i.e. its action takes us out of the domain of definition of the Hamiltonian, see [12],[13] for the above Ehrenfest theorem, and [14] for the related case of the extension of Hellmann-Feynman theorem). In what follows, we will develop a new methodology that extends even beyond the above Ehrenfest theorem (2) and applies mostly in the case of optical transitions by again involving the non-Hermitian boundary terms of eq. (2) and by generalizing even further to non-diagonal matrix elements (hence not only to the expectation values of (2)) and to left and right states that follow different Hamiltonians (but mutually related, their difference being an additive *t*-dependent perturbation term). Even more generally, our extended theorem can describe processes occurring in molecular orbitals as well as hoppings between different atoms or molecules.

## 2. The off-diagonal Ehrenfest theorem

We now generalize even further the above discussion: Suppose that we have a static Hamiltonian denoted by $H^0$, given by the following expression

$$H^0 = \frac{\left(\vec{p}+\frac{e}{c}\vec{A}(\vec{r})\right)^2}{2m} + V(\vec{r}),$$

with $\vec{A}(\vec{r})$ the magnetic vector potential and $V(\vec{r})$ a scalar potential energy with *e* being the electronic charge. At a given time *t*> 0, $H^0$ transforms into another, time-dependent Hamiltonian $H(t)$ by some mechanism (i.e. solar photon absorption) that can be interpreted as adiabatic perturbation of some parameter or by adding to $H^0$ an extra time-dependent perturbation term:

$$H(t) = H^0 + H'(\vec{r},t)$$

where $H'(\vec{r}, t)$ is a perturbative term that can be introduced through the position and the momentum operators always in position representation ($\vec{p} = -i\hbar\vec{\nabla}_r$). We suppose that the perturbation occurs at a specific instance $t=0$, so that we can write $H'(\vec{r}, t) = F(\vec{r}, t)\theta(t)$. Let $f(\vec{r}, t)$ and $\Psi(\vec{r}, t)$ be the most general solutions of the following Schrödinger equations:

$$H^0 f = i\hbar \frac{df}{dt} \text{ (6) and } H(t)\Psi = i\hbar \frac{d\Psi}{dt} \text{ (7)}$$

In what follows, we will deal with cases of optical transitions between the states $f$ and $\Psi$ when an optical matrix element can be represented by the inner product $\langle f|B_l|\Psi\rangle$, with $\vec{B}$ being generally a vector operator whose a certain Cartesian component is $B_l$; normally, it can be either the momentum or the position operator, but in general it can be any operator. The time-evolution of the matrix element then reads:

$$\frac{d}{dt}\langle f|B_l|\Psi\rangle = \left\langle \frac{\partial}{\partial t}f \middle| B_l \middle| \Psi \right\rangle + \left\langle f \middle| \frac{\partial}{\partial t} B_l \middle| \Psi \right\rangle + \left\langle f \middle| B_l \middle| \frac{\partial}{\partial t} \Psi \right\rangle = \left\langle f \middle| \frac{\partial}{\partial t} B_l \middle| \Psi \right\rangle + \frac{i}{\hbar}\langle H^0 f|B_l|\Psi\rangle - \frac{i}{\hbar}\langle f|B_l|H\Psi\rangle,$$
(8)

where we have made use of the above Schrödinger equations (eq. (6) and (7)). Next, we add and subtract the term $\frac{i}{\hbar}\langle f|B_l|H^0\Psi\rangle$ and make use of the fact that $\langle f|B_l|H^0\Psi\rangle = -\langle f|[H^0, B_l]|\Psi\rangle + \langle f|H^0 B_l|\Psi\rangle$ and that $H = H^0 + H'$ to find:

$$\frac{d}{dt}\langle f|B_l|\Psi\rangle = \left\langle f \middle| \frac{\partial}{\partial t} B_l \middle| \Psi \right\rangle + \frac{i}{\hbar}\langle f|[H^0, B_l]|\Psi\rangle - \frac{i}{\hbar}\langle f|B_l|H'\Psi\rangle + \frac{i}{\hbar}\langle H^0 f|B_l|\Psi\rangle - \frac{i}{\hbar}\langle f|H^0 B_l|\Psi\rangle \text{ (9)}$$

We now calculate the last two terms (which, generally, do not cancel out, due to the appearance of possible emergent non-Hermiticity of the kinetic energy of $H^0$, as seen in [1]), namely:

$$\langle H^0 f|B_l|\Psi\rangle - \langle f|H^0 B_l|\Psi\rangle = -\frac{\hbar^2}{2m}\langle \nabla^2 f|B_l|\Psi\rangle + \frac{\hbar^2}{2m}\langle f|\nabla^2 B_l|\Psi\rangle + \frac{i\hbar e}{mc}\langle \vec{A}.\vec{\nabla}f|B_l|\Psi\rangle + \frac{i\hbar e}{mc}\langle f|\vec{A}.\vec{\nabla}B_l|\Psi\rangle$$
(10)

so that, by using the Green's theorem (in a very similar manner as was proved in [1]) we conclude to the following equation that rigorously describes the dynamical development of the optical element:

$$\frac{d}{dt}\langle f|B_l|\Psi\rangle = \left\langle f \middle| \frac{\partial}{\partial t} B_l \middle| \Psi \right\rangle + \frac{i}{\hbar}\langle f|[H^0, B_l]|\Psi\rangle - \frac{i}{\hbar}\langle f|B_l|H'\Psi\rangle - \oint \vec{J}_{gen}^{f,\Psi}.d\vec{S}, \text{ (11)}$$

or, in differential form (using the above for any volume element of the system), we obtain the following extension of the continuity equation:

$$\vec{\nabla}.\vec{J}_{gen}^{f,\Psi} + \frac{d\rho_{gen}^{f,\Psi}}{dt} = f^*\left(\frac{\partial}{\partial t}B_l + \frac{i}{\hbar}[H^0, B_l] - \frac{i}{\hbar}B_l H'\right)\Psi \text{ (12)}$$

with $\vec{J}_{gen}^{f,\Psi} = \frac{i\hbar}{2m}\left(\vec{\nabla}f^* B_l\Psi - f^*\vec{\nabla}(B_l\Psi)\right) + \frac{e}{mc}\vec{A}f^* B_l\Psi$ (13) an off-diagonal generalization of flow density (which is an off-diagonal version of the previously mentioned $\vec{J}_{gen}$) and $\rho_{gen}^{f,\Psi} = f^* B_l\Psi$ (14) an off-diagonal generalization of the density of the Hermitian operator $B_l$. Eq. (11) can be viewed as a generalization of eq. (2) (to $f \neq \Psi$ and to $H' \neq 0$) and eq. (12) as a corresponding generalization of eq. (5). In the special case where $B_l = 1$, the identity operator, eq. (12) becomes:

$$\vec{\nabla}.\vec{J}_{gen}^{f,\Psi} + \frac{d\rho_{gen}^{f,\Psi}}{dt} = -\frac{i}{\hbar}f^* H'\Psi \text{ (15) with } \rho_{gen}^{f,\Psi} = f^*\Psi \text{ (16)}$$

A few comments are then worth making: The off-diagonal generalized density is actually the probability amplitude to make a transition from the initial (single-) eigenstate $f$ to the final (linear combination) state $\Psi$. Because the system (or the perturbation) is time-dependent, the off-diagonal generalized current density plays the role of the transition probability flow, i.e. there is a finite chance of the particle being energetically transported to a final state due to the action of the perturbation. The right hand-side of eq. (15) generates the transition probability in a type of local continuity equation, and acts as a source term. Eq. (11) gives another insight to the problem: the time flow of $\langle f|B_l|\Psi\rangle$ is governed by bulk terms and a surface term (the surface integral of $\vec{J}_{gen}^{f,\Psi}$) which gives one the opportunity to study a given problem from a dual perspective: the bulk physics and the surface physics, as we shall see below. It should be noted that eq. (12) is valid as written only if $\Psi$ is a solution of (7) (with Hamiltonian $H$) and $f$ is a solution of (6) (with Hamiltonian $H^0$).

At this point, to make sure that things are as clear as possible, we turn our attention to the solution $\Psi$; we reemphasize that this is a solution of $H(t)\Psi = i\hbar \frac{d\Psi}{dt}$, and is generally not connected with $f$, which is a solution of a different Schrödinger equation: $H^0 f = i\hbar \frac{df}{dt}$. In cases of solar absorption, it is reasonable to assume that $\Psi$ can be written as a linear combination of all $f$'s-eigenstates of $H^0$. Equation (12) is actually modified in form if one chooses to use - instead of $\Psi$- another eigenstate of $H^0$; Suppose i.e. that we are interested in the time evolution of the matrix element $\langle f_s|B_l|f_n\rangle$, with $f_s$, $f_n$ two orthogonal eigenstates of $H^0$ and $B_l$ is either the momentum or the position operator. In this case, after following a similar methodology as the one used to derive eq. (12), we arrive at the following generalized continuity equation:

$$\vec{\nabla}\cdot\vec{J}_{gen}^{s,n} + \frac{d\,\rho_{gen}^{s,n}}{dt} = f_s^*\left(\frac{\partial}{\partial t}B_l + \frac{i}{\hbar}[H^0, B_l]\right)f_n \quad \text{with} \quad \rho_{gen}^{s,n} = e^{\frac{i}{\hbar}(\varepsilon_s - \varepsilon_n)t} f_s^* f_n \quad (17)$$

which lacks - if compared to (12) - the perturbation term. This equation will also be used quite often in what follows.

### 3. Application of the generalized Ehrenfest theorem in cases with Fermi Golden Rule

To estimate the probability of an optical transition from the initial Quantum Mechanical state $f_l(\vec{r})$ (a single eigenfunction of $H^0$, solution of $H^0 f_l = \varepsilon_l f_l$), in which case $f = f_l(\vec{r})e^{-\frac{i}{\hbar}\varepsilon_l t}$, to the final state $\Psi = \sum_n a_n(t) f_n(\vec{r}) e^{-\frac{i}{\hbar}\varepsilon_n t}$ (a general solution of $H\Psi = i\hbar \frac{\partial \Psi}{\partial t}$) that can always be written as a linear combination of all $f_n(\vec{r})$ states of $H^0$ with time dependent coefficients $a_n(t)$, we must calculate the time evolution of $\langle f|1|\Psi\rangle$. Using eq. (15) we find:

$$\frac{d[f^*(\vec{r},t)\Psi(\vec{r},t)]}{dt} = -\vec{\nabla}\cdot\vec{J}_{gen}^{f,\Psi} - \frac{i}{\hbar}\sum_n e^{\frac{i}{\hbar}(\varepsilon_l - \varepsilon_n)t} a_n(t) H'_{l,n} \quad (18)$$

with $H'_{l,n} = f_l^* H' f_n$ and the $\vec{\nabla}\cdot\vec{J}_{gen}^{f,\Psi}$ term can be determined as follows: From eq. (13) and the definitions of $f$ and $\Psi$ we have that (always for $B_l$ = identity operator and assuming (for simplicity) that no vector potentials are present, as typical in solar cells)

$$\vec{J}_{gen}^{f,\Psi} = \frac{i\hbar}{2m}\left(\vec{\nabla}f^*\Psi - f^*\vec{\nabla}\Psi\right) = \frac{i\hbar}{2m}\sum_n a_n(t)\, e^{\frac{i}{\hbar}(\varepsilon_l-\varepsilon_n)t}\left(\vec{\nabla}f_l^*(\vec{r})f_n(\vec{r}) - f_l^*(\vec{r})\vec{\nabla}f_n(\vec{r})\right)$$

$$= \sum_n a_n(t)\, \vec{J}_{gen}^{l,n}$$

with $\vec{J}_{gen}^{l,n} = \frac{i\hbar}{2m} e^{\frac{i}{\hbar}(\varepsilon_l-\varepsilon_n)t}\left(\vec{\nabla}f_l^*(\vec{r})f_n(\vec{r}) - f_l^*(\vec{r})\vec{\nabla}f_n(\vec{r})\right)$. Now, according to eq. (17), and the fact that both $f_l$ and $f_n$ are eigenfunctions of $H^0$, we have that this generalized current obeys $\vec{\nabla}\cdot\vec{J}_{gen}^{l,n} + \frac{d\rho_{gen}^{l,n}}{dt} = 0$, with $\rho_{gen}^{l,n} = e^{\frac{i}{\hbar}(\varepsilon_l-\varepsilon_n)t} f_l^*(\vec{r})f_n(\vec{r})$ so that

$$\vec{\nabla}\cdot\vec{J}_{gen}^{f,\Psi} = \sum_n a_n(t)\vec{\nabla}\cdot\vec{J}_{gen}^{l,n} = -\sum_n a_n(t)\frac{d\rho_{gen}^{l,n}}{dt} = -\frac{i}{\hbar}\sum_n a_n(t)(\varepsilon_l-\varepsilon_n)e^{\frac{i}{\hbar}(\varepsilon_l-\varepsilon_n)t}f_l^*(\vec{r})f_n(\vec{r}) \quad (19)$$

with $\varepsilon_n$ the energy levels of the unperturbed Hamiltonian $H^0$. Integrating then eq. (18) over the whole volume of the system, we get:

$$\frac{d(\int d^3r\, f^*\Psi)}{dt} = \frac{da_l(t)}{dt}, \text{ and}$$

$$\frac{da_l(t)}{dt} = -\frac{i}{\hbar}\sum_n a_n(t)e^{\frac{i}{\hbar}(\varepsilon_l-\varepsilon_n)t}\left[(\varepsilon_n-\varepsilon_l)\int d^3r\, f_l^*f_n(\vec{r}) + \langle H'\rangle_{l,n}\right] = -\frac{i}{\hbar}\sum_n a_n(t)e^{\frac{i}{\hbar}(\varepsilon_l-\varepsilon_n)t}\langle H'\rangle_{l,n}$$
**(20)**

due to the orthonormality of $f_l(\vec{r})$ and $f_n(\vec{r})$. We observe that in this case, where the state $\Psi$ can be written as a linear combination of all independent eigenstates of the unperturbed Hamiltonian the integral of the $\vec{\nabla}\cdot\vec{J}_{gen}$ term vanishes. As a result, in the $B_l = 1$ case, the probability amplitude is not influenced by the appearance of boundary terms, but in a more general case, where the initial and the basis used for the expansion of the final states may not be orthogonal (e.g. transitions between different molecular orbitals), boundary terms will indeed be needed. The result (20) is the standard textbook result that leads to Fermi Golden Rule (after the usual approximations on the coefficients are made).

## 4. Application in Optical transitions

The off-diagonal non-Hermitian boundary terms may have potential consequences on the optical properties of semiconducting systems if one appropriately applies the extended Ehrenfest theorem for the optical matrix elements. In what follows, we will make use of eqs. (11), (12) and (17) to some quantum mechanical problems that are affected by a time-dependent perturbation (i.e. a solar photon absorption) and calculate the dipole matrix element and the momentum matrix element (in case that we have a scalar or a vector potential to describe the interaction of matter with the electric field of light). The interaction term *H' will be set to zero, because we will only be interested in single eigenstates of the unperturbed Hamiltonian $H^0$ as the initial and final states* (cf. eq. (17)). Let us start from the time-dependence of the transition dipole matrix element $\langle f|\vec{r}|\Psi\rangle$ with $f(\vec{r},t) = f_s(\vec{r})e^{-\frac{i}{\hbar}\varepsilon_s t}$ and $\Psi(\vec{r},t) = f_n(\vec{r})e^{-\frac{i}{\hbar}\varepsilon_n t}$, both solutions of $H^0$. The usefulness of our results is demonstrated in the following way: For single eigenstates as the example above, and time-independent operators, we have that the following relation holds:

$$\frac{d}{dt}\langle f|B_l|\Psi\rangle = i\omega_{s,n}\langle f_s|B_l|f_n\rangle \quad (21) \quad \text{with } \omega_{s,n} = \frac{\varepsilon_s-\varepsilon_n}{\hbar}$$

i.e. the off-diagonal matrix element between single eigenkets of any physical observable is proportional to its time-derivative. In contrast, when the diagonal matrix element is used instead, the time derivative gives a null result, as indeed expected, because the time-phase factor of the single eigenkets is eliminated. This result gives one the potential to express the optical transition element in terms of its time derivative (simplifying, as we shall see, in many cases the calculation load). We proceed with three simple but important examples.

### 4.1. Free particle in 1 and 2 dimensions

It is convenient to first present a simple example in 2D: Consider a particle in the interior of a 2D rectangle ($L_x \times L_y$) with vanishing vector and scalar potentials $(\vec{A}, V) = 0$ and periodic boundary conditions along the sides $L_x$ and $L_y$ : In this free particle case, the normalized eigenfunctions of the Hamiltonian are:

$$f_{\vec{k}}(\vec{r}) = \frac{1}{\sqrt{L_x L_y}} e^{i\vec{k}\cdot\vec{r}}, \text{ (22)}$$ with $k_x = 2\pi \frac{n_x}{L_x}, k_y = 2\pi \frac{n_y}{L_y}$ **(23)**, $n_x, n_y = 0, \pm 1, \pm 2, \ldots$ with eigenenergies:

$$\varepsilon_{\vec{k}}^0 = \frac{\hbar^2 (k_x^2 + k_y^2)}{2m} \text{ (24)}$$

Let us then consider the $x$-component of the position as our input operator in eq. (17) to find (with $f = \langle \vec{r} | \vec{k}' \rangle$ and $\Psi = \langle \vec{r} | \vec{k} \rangle$ being two different, orthonormal eigenfunctions):

$$\frac{d}{dt}\langle x \rangle_{\vec{k}, \vec{k}'} = i\omega_{\vec{k}, \vec{k}'} \langle x \rangle_{\vec{k}, \vec{k}'} = \frac{\langle p_x \rangle_{\vec{k}, \vec{k}'}}{m} - \oint \vec{J}_{gen_\perp}^{\vec{k}, \vec{k}'} dl, \text{ (25)}$$

with $\langle p_x \rangle_{\vec{k}, \vec{k}'} = 0$ (always for $\vec{k} \neq \vec{k}'$), and the line integral along the line boundary is the integral of the transverse component of $\vec{J}_{gen}^{\vec{k}, \vec{k}'}$ ($\vec{J}_{gen_\perp}^{\vec{k}, \vec{k}'}$) to the edges of the rectangle, namely:

$$\oint \vec{J}_{gen_\perp}^{\vec{k}, \vec{k}'} dl = \int_0^{L_x} \vec{J}_{gen,y}^{\vec{k}, \vec{k}'}(x, 0) dx - \int_0^{L_x} \vec{J}_{gen,y}^{\vec{k}, \vec{k}'}(x, L_y) dx - \int_0^{L_y} \vec{J}_{gen,x}^{\vec{k}, \vec{k}'}(0, y) dy + \int_0^{L_y} \vec{J}_{gen,x}^{\vec{k}, \vec{k}'}(L_x, y) dy =$$
$$\frac{\hbar}{2m} e^{i\omega_{\vec{k}, \vec{k}'} t} \frac{1}{L_x L_y} (k_y' + k_y) \int_0^{L_x} x e^{i(k_x - k_x')x} dx - \frac{\hbar}{2m} e^{i\omega_{\vec{k}, \vec{k}'} t} \frac{1}{L_x L_y} (k_y' + k_y) \int_0^{L_x} x e^{i(k_x - k_x')x} dx - 0 + 0 = 0 \text{ (26)}$$

where we used $\vec{J}_x^{\vec{k}, \vec{k}'}(x, y) = \frac{i\hbar}{2m} e^{i\omega_{\vec{k}, \vec{k}'} t} \left( \frac{\partial f_{\vec{k}'}^*}{\partial x} x f_{\vec{k}} - f_{\vec{k}'}^* f_{\vec{k}} - x f_{\vec{k}'}^* \frac{\partial f_{\vec{k}}}{\partial x} \right)$, $\vec{J}_y^{\vec{k}, \vec{k}'}(x, y) = \frac{i\hbar}{2m} e^{i\omega_{\vec{k}, \vec{k}'} t} \left( \frac{\partial f_{\vec{k}'}^*}{\partial y} x f_{\vec{k}} - x f_{\vec{k}'}^* \frac{\partial f_{\vec{k}}}{\partial y} \right)$ and $\omega_{\vec{k}, \vec{k}'} = \frac{\varepsilon_{\vec{k}'}^0 - \varepsilon_{\vec{k}}^0}{\hbar}$.

If we put everything into eq. (25), the final result seems to be:

$$\frac{d}{dt}\langle x \rangle_{\vec{k}, \vec{k}'} = 0, \text{ for } \vec{k} \neq \vec{k}' \text{ (27)}$$

To independently check the validity of this result, we straightforwardly proceed with the verification of eq. (27) (always for $\vec{k} \neq \vec{k}'$):

$$\frac{d}{dt}\langle x \rangle_{\vec{k}, \vec{k}'} = \frac{i\omega_{\vec{k}, \vec{k}'}}{L_x L_y} e^{i\omega_{\vec{k}, \vec{k}'} t} \int_0^{L_x} dx\, x e^{i(k_x - k_x')x} \int_0^{L_y} dy\, e^{i(k_y - k_y')y}$$
$$= \frac{i\omega_{\vec{k}, \vec{k}'}}{L_x L_y} e^{i\omega_{\vec{k}, \vec{k}'} t} \int_0^{L_x} dx\, x e^{i(k_x - k_x')x} \left[ \frac{e^{i(k_y - k_y')L_y} - 1}{i(k_y - k_y')} \right] = 0$$

(the last vanishing due to eq. (23)). In two dimensions we get a null result, but it is interesting to note that if we had carried out our calculations in the case of 1D (an electron in a linear region of length $L_x$, with periodic boundary conditions), namely:

$f_k(x) = \frac{1}{\sqrt{L_x}} e^{ikx}$, with $k_x = 2\pi \frac{n_x}{L_x}$, and $\varepsilon_k^0 = \frac{\hbar^2 k^2}{2m}$ we would have found that (after applying the integrated version of eq. (17)):

$$\frac{d}{dt}\langle x \rangle_{k,k'} = \frac{\langle p_x \rangle_{k,k'}}{m} - \vec{J}_{gen}^{k,k'}\Big|_0^{L_x}$$

with $\langle p_x \rangle_{k,k'} = 0$ (always for $\vec{k} \neq \vec{k'}$) and

$$\vec{J}_{gen}^{k,k'}\Big|_0^{L_x} = \frac{i\hbar}{2mL_x} e^{i\omega_{\vec{k},\vec{k'}}t} e^{i(k-k')x}(-1 - i(k+k')x)\Big|_0^{L_x} = \frac{\hbar\pi}{mL_x} e^{i\omega_{\vec{k},\vec{k'}}t}(n_x + n_x')$$

so that $\frac{d}{dt}\langle x \rangle_{k,k'} = -\frac{\hbar\pi}{mL_x}(n_x + n_x') e^{i\omega_{\vec{k},\vec{k'}}t}$, hence a non-vanishing result.

This can again be verified for the case of $k \neq k'$, by straightforwardly calculating the time derivative, namely:

$$\frac{d}{dt}\langle x \rangle_{k,k'} = \frac{1}{L_x} e^{i\omega_{\vec{k},\vec{k'}}t} \int_0^{L_x} dx\, x e^{i(k-k')x} = \frac{i}{L_x} \omega_{\vec{k},\vec{k'}} e^{i\omega_{\vec{k},\vec{k'}}t} \int_0^{L_x} dx\, x e^{i(k-k')x} =$$

$$\frac{i}{L_x} \omega_{\vec{k},\vec{k'}} e^{i\omega_{\vec{k},\vec{k'}}t} x \frac{e^{i(k-k')x}}{i(k-k')}\Big|_0^{L_x} = \frac{1}{(k-k')} \omega_{\vec{k},\vec{k'}} e^{i\omega_{\vec{k},\vec{k'}}t} = -\frac{\hbar\pi}{mL}(n_x + n_x') e^{i\omega_{\vec{k},\vec{k'}}t} \quad (28)$$

One can also add here the expected $\frac{d}{dt}\langle x \rangle_{k,k'} = 0$ in the case of $k = k'$ (as the expectation value in a single-eigenstate is indeed $t$-independent, meaning that its time-derivative will result to zero) that actually motivated the discussion in [1] and was developed there in full detail.

We notice from the above example that the dimensionality of a given problem is very important because the extra spatial degrees of freedom may affect the photon absorbance differently; for the previous 1D case, transition probabilities between different wave-number states are possible, as given by eq. (28). In the 2D case however, this is not always possible (as predicted by eq. (27)) for a linearly polarized electric field. Transitions are possible, however, in the case of circularly polarized (or any directionally time-varying) electric field, in which case the above electric dipole element may not vanish. Similar conclusions can be also drawn for the 3D case.

### 4.2. Quantum bouncing ball

After the previous example, viewed as a preliminary step, we now turn our attention to the off-diagonal momentum optical matrix element $\langle \vec{\Pi} \rangle_{i,n}$, with $\vec{\Pi} = \vec{p} + \frac{e}{c}\vec{A}$ the kinetic momentum. Using for simplicity a specific component of the vector operator $\vec{\Pi}$, i.e. $B_l = \Pi_x = p_x + \frac{e}{c}A_x$ (and for the case $H' = 0$) in eq. (17) we arrive at the following equation:

$$\vec{\nabla}.\vec{J}_{gen}^{f,\Psi} + \frac{d\rho_{gen}^{f,\Psi}}{dt} = f^*\left(-\frac{eB_z}{mc}\Pi_y + \frac{eB_y}{mc}\Pi_z - \frac{\partial V}{\partial x}\right)\Psi \quad (29)$$

Then, for $f$ and $\Psi$ regarded as two distinct eigenfunctions of the same Hamiltonian (hence, there is no need to include $H'$, as explained earlier)), and by using (through eq. (21))

$$\frac{d\,\rho_{gen}^{f,\Psi}}{dt} = \frac{d}{dt}\left(e^{i\frac{(\varepsilon_f-\varepsilon_\psi)t}{\hbar}}f^*(\vec{r})\Pi_x\Psi(\vec{r})\right) = i\omega_{f,\psi}\rho_{gen}^{f,\Psi}(\vec{r}) \quad (30)$$

with $\omega_{f,\psi} = \frac{(\varepsilon_f-\varepsilon_\psi)}{\hbar}$, and by also defining the cyclotron frequencies $\omega_i = eB_i/mc$ with $i = x,y,z$, we obtain from the integrated version of (29) (and always for the choice $B_l = \Pi_x = p_x + \frac{e}{c}A_x$) that

$$i\omega_{f,\psi}\langle\Pi_x\rangle_{f,\psi} = -\omega_z\langle\Pi_y\rangle_{f,\psi} + \omega_y\langle\Pi_z\rangle_{f,\psi} - \langle\frac{\partial V}{\partial x}\rangle_{f,\psi} - \oint \vec{J}_{gen}^{f,\Psi}\cdot d\vec{S} \quad (31)$$

with $\vec{J}_{gen}^{f,\Psi}$ given by eq. (13) with $B_l = \Pi_x$. In [2], the authors use the Ehrenfest theorem to calculate $\langle\Pi_x\rangle_{f,\psi}$ neglecting the boundary terms, which terms however can in principle be very important and can contribute equally to the overall result. The above equation (31) has the advantage that, it can relate the optical element $\langle\Pi_i\rangle_{f,\psi}$ with the effective bulk force acting on the particle and with a boundary term (that can be interpreted as a surface force) as a result of the interaction with the electromagnetic field.

To underline the important physical consequences of the above new non-Hermitian terms, we will calculate the last term appearing in (31) for the case of an electron in a triangular well (described by a homogeneous electric field $E$) in 1D without any magnetic fields or vector potentials present. In this case, the wavefunctions are represented by the Airy functions: $\Psi_n = C_n Ai[(x - \varepsilon_n/eE)/l_f]$, with $C_n$ a normalization constant, $Ai[x]$ the Airy functions, $\varepsilon_n$ the energy levels and $l_f = (\hbar^2/2meE)^{1/3}$. Let $f = C_{n'}Ai[(x - \varepsilon_{n'}/eE)/l_f]$ and $\Psi = C_n Ai[(x - \varepsilon_n/eE)/l_f]$ be two different, linearly independent solutions of the Schrödinger equation:

$$\Psi'' - \frac{2meE}{\hbar^2}\left(x - \frac{\varepsilon_n}{eE}\right)\Psi = 0 \quad (32)$$

The boundary condition at $x=0$ allows us to directly relate the energy eigenvalues $\varepsilon_n$ to the roots $a_n$ of the Airy function, namely

$$Ai\left[-\varepsilon_n/eEl_f\right] = 0 \Rightarrow \varepsilon_n = -eEl_f a_n \text{ with } n=1,2... \quad (33)$$

The generalized current density (eq. (13)) with $B_l = \Pi_x = p_x = -\frac{i\hbar\partial}{\partial x}$ (since now $A=0$)) then reads

$$\vec{J}_{gen}^{f,\Psi} = e^{i\omega_{n,n'}t}\frac{\hbar^2}{2m}C_{n'}^*C_n\left(Ai^{*\prime}\left[\frac{x}{l_f}+a_{n'}\right]Ai'\left[\frac{x}{l_f}+a_n\right] - Ai^*\left[\frac{x}{l_f}+a_{n'}\right]Ai''\left[\frac{x}{l_f}+a_n\right]\right) \quad (34)$$

with $\omega_{n,n'} = \frac{\varepsilon_{n'}-\varepsilon_n}{\hbar}$. Because the Airy function $Ai\left[\frac{x}{l_f}+a_{n'}\right]$ (and its derivative) at the asymptotic limit $x\to\infty$ vanishes, and so does at $x=0$ (due to the infinite potential wall), we find that the only surviving boundary term (the last flux term of eq. (31) in this 1D case) is the product of derivatives at $x=0$, hence the final term of (31) has the form

$$\vec{J}_{gen}^{f,\Psi}\Big|_0^\infty = -\frac{\hbar^2}{2ml_f^2}C_{n'}^*C_n e^{i\omega_{n,n'}t}(Ai^{*\prime}[a_{n'}]Ai'[a_n])\ (35)$$

By then using the normalization constant $C_n = 1/[\sqrt{l_f}Ai'(a_n)]$ we find that the non-Hermitian boundary term in this problem is simply

$$\vec{J}_{gen}^{f,\Psi}\Big|_0^\infty = -\frac{\hbar^2}{2ml_f^3}e^{i\omega_{n,n'}t}\ (36)$$

Now, considering that $\langle\Pi_y\rangle_{n,n'} = 0$, $\langle\Pi_z\rangle_{n,n'} = 0$ (for a 1D case) and that $\frac{\partial V}{\partial x} = eE$ (the bulk force, which is homogeneous) it is immediate that $\langle\frac{\partial V}{\partial x}\rangle_{n,n'} = 0$, due to the orthogonality of $n$ and $n'$ (in eq. (31)) and we therefore note that $\langle\Pi_x\rangle_{n,n'}$ is proportional to the boundary term. On the other hand, we should point out that the full potential profile consists of both bulk and boundary terms, namely:

$$V = \lim_{V_0\to\infty} V_0\theta(-x) + eEx\theta(x)\ (37)$$

with $\theta(x)$ the Heaviside step function, so that the full force equation should be related to $\frac{\partial V}{\partial x} = \lim_{V_0\to\infty} V_0\delta(x) + eE\theta(x)$, with $\delta(x)$ the Dirac delta function. But, in doing so, there is a danger of double-counting the force contribution. What the integrated version of (29) (or eq. (31)) actually succeeds in doing is to divide the problem into a bulk term and a surface term, which can be treated separately. To correctly calculate $\langle\frac{\partial V}{\partial x}\rangle_{n,n'}$ in eq. (29) we only need to use the bulk force element, $eE\theta(x)$, which actually gives a null result. All surface terms (here forces, momentum transfer etc.) are automatically built-in the last term of eq. (31), and no further calculations to determine the wavefunction are needed. In our case we obtain

$$\frac{d\langle p_x\rangle_{n,n'}}{dt} = i\omega_{f,\psi}\langle p_x\rangle_{n,n'} \Rightarrow \langle p_x\rangle_{n,n'} = \frac{-i\hbar^2}{2ml_f^3\omega_{n,n'}}e^{i\omega_{n,n'}t}\ (38)$$

and we can see that eqs (38) and (36) are indeed consistent with eq. (31) with the vanishing of $\langle\frac{\partial V}{\partial x}\rangle_{n,n'}$. [Further discussion on this is given at the end of this Section.] It should also be noted that, in spite of the claim in [2], it is possible to find a way to analytically show the above result independently, with use of Airy function properties, and this is presented in Appendix 1. Furthermore, for completeness, we here carry out corresponding calculations, but now using the position operator as the input operator ($B_l = x$) in the integral version of eq. (17) (we do this for comparison purposes - assuming that the electromagnetic field is now coupled through a dipole moment interaction term). In this case we have

$$i\omega_{n',n}\langle\Psi_{n'}|x|\Psi_n\rangle = \frac{1}{m}\langle\Psi_{n'}|p_x|\Psi_n\rangle + \vec{J}_{gen}^{f,\Psi}\Big|_0^\infty,$$

with $\langle\Psi_{n'}|p_x|\Psi_n\rangle = \frac{-i\hbar^2}{2ml_f^3\omega_{n,n'}}e^{i\omega_{n,n'}t}$ as given by eq. (38) and

$$\vec{J}_{gen}^{f,\Psi}\Big|_0^\infty = \frac{i\hbar}{2m}\Psi'_{n'}x\Psi_n - \Psi_{n'}(\Psi_n + x\Psi'_n)\Big|_0^\infty = 0$$

So that we get the expected result (that demonstrates the duality between the choices of the momentum and position to describe the electromagnetic radiation), namely:

$$\langle \Psi_{n'} | x | \Psi_n \rangle = \frac{\hbar^2}{2ml_f^3 \omega_{n,n'}^2} e^{i\omega_{n,n'}t}$$

It is important to notice that in higher dimensionality cases, where line and surface integrals of the non-Hermitian terms appear, the boundary terms might not be zero; in these cases a more careful calculation is necessary (and this deserves to be the focus of future work).

### 4.3. Particle in an infinite potential well

Let us finally see a simpler example from elementary Quantum Mechanics (now without electric field) where the new surface terms proposed here might again be important: Consider an electron in a 1D quantum potential well with infinite walls and $(\vec{A}, V) = 0$ inside the cell, with normalized eigenfunctions

$$\Psi_n = \sqrt{\frac{2}{d}} \sin\left(\frac{n\pi x}{d}\right) e^{-\frac{i\varepsilon_n t}{\hbar}} \quad (39)$$

and eigenenergies $\varepsilon_n = \frac{\hbar^2 \pi^2 n^2}{2md^2}$ with $n = 1, 2..$ and $d$ is the quantum well's length. We have that $\frac{\partial \Psi}{\partial x} = \sqrt{\frac{2}{d}} \frac{n\pi}{d} \cos\left(\frac{n\pi x}{d}\right)$ and $\frac{\partial^2 \Psi}{\partial x^2} = -\sqrt{\frac{2}{d}} \left(\frac{n\pi}{d}\right)^2 \cos\left(\frac{n\pi x}{d}\right)$, so that for $l \neq n$ (and with momentum $B_l = p_x$ as input operator) we obtain the boundary flux term

$$\vec{J}_{gen}^{l,n}\Big|_0^d = \frac{i\hbar}{2m} [\vec{\nabla}\Psi_l^* p_x \Psi_n - \Psi_l^* \vec{\nabla}(p_x \Psi_n)]\Big|_0^d = \frac{i\hbar}{2m}\left[-i\hbar \frac{2n\pi}{d^2} \frac{l\pi}{d} \cos\left(\frac{n\pi x}{d}\right) \cos\left(\frac{l\pi x}{d}\right)\right]\Big|_0^d e^{i\omega_{l,n}t} =$$
$$\hbar^2 \frac{nl\pi^2}{md^3} [\cos(n\pi)\cos(l\pi) - 1] e^{i\omega_{l,n}t} \quad (40)$$

so that by using the integral form of eq. (17) (which is the 1D version of eq. (31)) we find (after using $\frac{\partial V}{\partial x} = 0$) that

$$\frac{d}{dt}\langle p_x \rangle_{l,n} = -\hbar^2 \frac{nl\pi^2}{md^3} [\cos(n\pi)\cos(l\pi) - 1] e^{i\omega_{l,n}t} \quad (41)$$

In (41), only terms that satisfy the condition $n-l$ =odd survive, and they give:

$$\frac{d}{dt}\langle p \rangle_{l,n} = -2i\hbar^2 \frac{nl\pi^2}{md^3 \omega_{l,n}} e^{i\omega_{l,n}t} \quad (42)$$

For the case $l = n$, we also have that $\vec{J}_{gen}^{l,l} = 0$ and $\frac{d}{dt}\langle p \rangle_{l,l} = 0$.

We therefore conclude that the non-Hermitian terms are of vital importance when it comes to calculate the time dependence of optical matrix elements, and should always be included. Generally, we can see from eq. (31) that, if the surface term were indeed zero, we could write directly a certain component of the optical matrix element $\langle \Pi_i \rangle_{f,\psi}$ as a function of the 'effective force', $\langle \vec{\nabla} V \rangle_{f,\psi}$ and simplify the calculations as already been done in [2]. However, this is not always the case, as the non-Hermitian terms appear as a consequence of a generalized conservation theorem. This is demonstrated more clearly in the comments that follow.

If, again, the momentum $p_x$ is chosen as an input operator in eq. (31), there is a bulk force contribution from the gradient of the potential and a surface contribution from the non-Hermitian term. While the potential gradient refers to the (off-diagonal) bulk force acting on the particle, the surface term incorporates the surface force directly (which is generally proportional to the product of the derivatives of the two transverse wavefunctions-as also seen earlier in example (35) and as will also be demonstrated below). Let us present another simple example**:** Consider once again the simple case of an electron in a 1D quantum well of length $L$ (with no-vector potential) and equation (31) in 1D, namely

$$i\omega_{f,\psi}\langle p_x\rangle_{f,\psi} = -\langle\frac{\partial V}{\partial x}\rangle_{f,\psi} - \vec{J}^{f,\Psi}_{gen}\Big|_0^L \quad (43)$$

with $\vec{J}^{f,\Psi}_{gen} = \frac{i\hbar}{2m}\left(\frac{\partial}{\partial x}f^* p_x\Psi - f^*\frac{\partial}{\partial x}(p_x\Psi)\right) = \frac{\hbar^2}{2m}\left(\frac{\partial}{\partial x}f^*\frac{\partial}{\partial x}\Psi - f^*\frac{\partial^2}{\partial x^2}\Psi\right)$ **(44).** In this case, the potential profile reads:

$$V = \lim_{V_0\to\infty} V_0[\theta(-x) + \theta(x-L)] \quad (45)$$

The gradient, $\frac{\partial V}{\partial x}$ reads: $\frac{\partial V}{\partial x} = \lim_{V_0\to\infty} V_0[\delta(-x) + \delta(x-L)]$ **(46)**. This is the surface force operator acting on the particle, while the bulk force operator is obviously $\frac{\partial V}{\partial x} = 0$. Eq. (43) can then be utilized in a twofold manner: If one chooses to use the boundary terms as in eq. (43), then, because the surface force information is already included in $\vec{J}^{f,\Psi}_{gen}\Big|_0^L$, (which is actually proportional to the product of the derivatives of the wavefunctions, because the second term in eq. (44) vanishes, and only the first term survives) and $\frac{\partial V}{\partial x}$ must only be the bulk force which is zero. On the other hand, if one wishes to neglect the boundary terms in (43), then the full potential profile (eq. (46)) must be used. Extra care is therefore needed in order to avoid double counting of the force contribution to $\langle p_x\rangle_{f,\psi}$. As a byproduct we note that, if in a problem there are periodic boundary conditions, then, the full version of eq. (43) must be used, because of the difficulty in obtaining the exact form of the surface potential.

Finally, we should re-emphasize that the generality of our results permits one to conduct calculations beyond the stationary states and consider delocalized states or even states that belong to different Hamiltonians, which is why states *f* and *Ψ* are left intentionally unspecified.

A final observation is worth making at the end of this Section and on a quite different matter. The example above, with the momentum as the input operator, is similar to the one used by Kellendonk [10], where a corresponding boundary term (but a diagonal element, that turned out to be the average force from the boundary) was shown to be *quantized* (and equal to the integrated density of states) – although the non-Hermitian (and emerging) nature of this quantized term was never pointed out. Such a quantized nature of a non-Hermitian boundary term had also come out in the Aharonov-Bohm ring problem in [1] with the position as the input operator, although the origin of that quantization was different (namely the standard Bohr type of quantization, when an integer number of wavelengths has to fit the circumference of the ring). Note also that our own (non-diagonal) result eq. (42) is also quantized, with a similar type of conventional quantization. It is then natural to speculate as to whether the observed quantizations might reflect a more general (universal) property of non-Hermitian boundary contributions**.** And although a general investigation is beyond the scope of the present article, we here provide one more result (non-Hermitian boundary contribution) for another example, this time a topologically nontrivial system, again with the position being the input operator, but in higher dimensionality**:** it is the key system in the area of the Integer Quantum Hall Effect [6] (a well-known system with topologically nontrivial states, now recognized as the first example of a topological insulator [7,8], with the well-known very special edge states**).** Our toy example is the Landau problem of noninteracting electrons moving in a macroscopic rectangle in the presence of a perpendicular homogeneous magnetic field, that can be directly solved in a Landau gauge, so that we have periodic boundary conditions along one Cartesian direction and open ends along the orthogonal one (these open ends being the edges of this 2D system). From the exact analytical solution of this Landau problem with an additional voltage applied to connect the two edges (hence in the additional presence of a uniform electric field *E* perpendicular to the edges) it turns out that there is a non-Hermitian term originating with the direction of periodicity that is quite similar to the one noted for the quantum ring in [1]**:** it is such that it has to cancel the usual velocity term (which is in turn equal to the global probability current along the same direction), due to the fact that the Ehrenfest theorem with input the position operator (and with respect to a single stationary state) should give a time-independent expectation value, hence its derivative is zero, hence the aforementioned cancellation. Taking then into account many non-interacting electrons that can be accommodated in *v* completely filled Landau Levels, it turns out that the total non-Hermitian contribution to the (transverse to *E*) electric current is quantized, with a value equal to $ve^2E/h$. But once again, although this gives another example of a quantized non-Hermitian term, it is obvious that a general investigation should clearly be based on a future article – dedicated to the issue of general conditions under which quantization of emergent non-Hermiticities may be expected.

## 5. Conclusions

We have shown how to apply an extended form of the Ehrenfest theorem in the case of optical matrix elements calculations. This generalized form, including non-Hermitian boundary terms defines a continuity equation describing the flow of a specific optical matrix element $\langle f_l|B_l|f_n\rangle$, with $B_l$ a Cartesian component of a Hermitian operator of the system, and it has been applied to some elementary quantum mechanical problems demonstrating the necessity of inclusion of the non-Hermitian terms. Dimensionality and given boundary conditions are crucial in the determination of certain generalized boundary currents, which separate the problem from its bulk response. For example, in some problems, the quantum force originates completely from the boundaries, while, in other problems, both the boundary and the bulk play equal roles. The applications in this paper with the separation into a bulk and a boundary contribution are an encouraging sign for a future successful applicability of the extended Ehrenfest theorem to topologically nontrivial quantum systems where bulk-boundary correspondence has been observed, such as the various Quantum Hall effects [6] either in conventional semiconductors or in pseudorelativistic (Dirac) systems such as Graphene, and even in more general topological materials such as Topological Insulators [7,8] or 3D Dirac and Weyl semimetals [9]. In those cases, and from an example on a Quantum Hall system provided in the last Section, one would expect the non-Hermitian boundary terms (manifestation of the topological anomalies [11]) to be quantized, also in accordance with few results in the literature [10] that however never mention the non-Hermitian (and emerging) nature of the quantized terms. This general expectation, however, needs to be verified through further study.

## Appendix 1.

We here prove the result (38) analytically:

$$\langle p_x\rangle_{n,n\prime} = -i\hbar \int_0^\infty dx\, \Psi_{n\prime}^* \frac{\partial \Psi_n}{\partial x} = \frac{-i\hbar^2}{2ml_f^3 \omega_{n,n\prime}} e^{i\omega_{n,n\prime}t}$$

Let $\Psi_{n'}^* = C_{n'} Ai[(x - \varepsilon_{n'}/eE)/l_f] e^{\frac{i\varepsilon_{n'}t}{\hbar}}$ and $\Psi_n^* = C_n Ai[(x - \varepsilon_n/eE)/l_f] e^{-\frac{i\varepsilon_n t}{\hbar}}$ be two orthogonal solutions of the Schrödinger equation:

$$\Psi_{n'}'' - \frac{2meE}{\hbar^2}\left(x - \frac{\varepsilon_{n'}}{eE}\right)\Psi_{n'} = 0 \quad (A1)$$

$$\Psi_n'' - \frac{2meE}{\hbar^2}\left(x - \frac{\varepsilon_n}{eE}\right)\Psi_n = 0 \quad (A2)$$

Multiply (A1) with $\Psi_n'$ and (A2) with $\Psi_{n'}'$, and add them up to find:

$$\Psi_{n'}''\Psi_n' + \Psi_n''\Psi_{n'}' - \frac{2meE}{\hbar^2}\left(x\Psi_{n'}\Psi_n' - \frac{\varepsilon_{n'}}{eE}\Psi_{n'}\Psi_n' + x\Psi_n\Psi_{n'}' - \frac{\varepsilon_n}{eE}\Psi_n\Psi_{n'}'\right) = 0 \quad (A3)$$

Integrate eq. (A3) with respect to x:

$$\int_0^\infty dx \left[\Psi_{n'}''\Psi_n' + \Psi_n''\Psi_{n'}' - \frac{2meE}{\hbar^2}\left(x\Psi_{n'}\Psi_n' - \frac{\varepsilon_{n'}}{eE}\Psi_{n'}\Psi_n' + x\Psi_n\Psi_{n'}' - \frac{\varepsilon_n}{eE}\Psi_n\Psi_{n'}'\right)\right] = 0 \quad (A4)$$

We will now make use of the properties of the Airy functions to simplify the results:

$$\int_0^\infty dx\, \Psi_{n'}''\Psi_n' = \Psi_{n'}'\Psi_n'\Big|_0^\infty - \int_0^\infty dx\, \Psi_{n'}'\Psi_n'' \quad (A5)$$

$$\int_0^\infty dx\, x\Psi_{n'}\Psi_n' = x\Psi_{n'}\Psi_n\Big|_0^\infty - \int_0^\infty dx\, x\Psi_n\Psi_{n'}' = -\int_0^\infty dx\, x\Psi_n\Psi_{n'}' \quad (A6)$$

$$\int_0^\infty dx\, \Psi_{n'}\Psi_n' = -\int_0^\infty dx\, \Psi_n\Psi_{n'}' \quad (A7)$$

Substituting (A5), (A6) and (A7) into (A4) we conclude to:

$$\Psi_{n'}'\Psi_n'\Big|_0^\infty - \frac{2m}{\hbar^2}(\varepsilon_n - \varepsilon_{n'})\int_0^\infty dx\, (\Psi_{n'}\Psi_n') = 0 \quad (A8)$$

Because $\langle p_x \rangle_{n,n'} = -i\hbar \int_0^\infty dx\, \Psi_{n'}^* \frac{\partial \Psi_n}{\partial x}$ we get: $-i\hbar \int_0^\infty dx\, (\Psi_{n'}\Psi_n') = -i\frac{\hbar^2 \Psi_{n'}'\Psi_n'\big|_0^\infty}{2m\omega_{n,n'}}$. Note that in the asymptotic limit $x \to \infty$, both the wavefunction and its derivative vanish, so that $\Psi_{n'}'\Psi_n'\big|_0^\infty = -\Psi_{n'}'(0)\Psi_n'(0) = -\frac{1}{l_f^3}$ and therefore

$$\langle p_x \rangle_{n,n'} = -i\frac{\hbar^2}{2m\omega_{n,n'}l_f^3} e^{i\omega_{n,n'}t} \quad (A9)$$

which coincides with eq. (38) that has been derived with much less effort in the main text, with use of the non-Hermitian boundary terms.

## 6. References

[1] G. Konstantinou, K. Kyriakou, K. Moulopoulos, "Emergent Non-Hermitian Contributions to the Ehrenfest and Hellmann-Feynman Theorems", International Journal of Engineering Innovations and Research 5(4), 248 (2016)

[2] A. G. Petrov and A. Shik, "Interlevel optical transitions in quantum wells", Physical Review B 48, 11883 (1993)

[3] H. R. Lewis Jr. and W. B. Riesenfeld, "An Exact Quantum Theory of the Time-Dependent Harmonic Oscillator and of a Charged Particle in a Time-Dependent Electromagnetic Field", J. Math. Phys. 10, 1458 (1969)

[4] R. N. Hill, "A Paradox Involving the Quantum Mechanical Ehrenfest Theorem", Am. J. Phys. 41, 736 (1973)

[5] F. M. Fernández and E. A. Castro, "Lecture Notes in Chemistry: Hypervirial Theorems", Springer-Verlag (1987)

[6] G. M. Graf, "Aspects of the Integer Quantum Hall Effect", Pure Mathematics: Volume 76, part 1 p.429 (2007)

[7] M. Z. Hasan, C. L. Kane, Colloquium: "Topological Insulators", Rev. Mod. Phys. 82, 3045 (2010)

[8] X.-L. Qi & S.-C. Zhang, "Topological insulators and superconductors", Rev. Mod. Phys. 83, 1057 (2011)

[9] X. Wan, A. M. Turner, A. Vishwanath and S. Y. Savrasov, "Topological semimetal and Fermi-arc surface states in the electronic structure of pyrochlore iridates", Phys. Rev. B 83, 205101 (2011)

[10] J. Kellendonk, "Topological quantization of boundary forces and the integrated density of states", arXiv:cond-mat/0311187

[11] V. Aldaya, M. Calixto and J. Guerrero, "Algebraic versus Topologic Anomalies", VI International Conference on Differential Geometry and Applications, Volume: Conference Proceedings of Masaryk University, Brno, pp. 495-502 (1996) ; [arXiv:hep-th/9702069]

[12] J. G. Esteve , "Anomalies in conservation laws in the Hamiltonian formalism", Phys. Rev. D 34, 674(R) (1986)

[13] J. G. Esteve, "Origin of the anomalies: The modified Heisenberg equation", Phys. Rev. D 66, 125013 (2002)

[14] J. G. Esteve et al, Phys. Lett. A 374(6) (2009) ; [arXiv:0912.4153]